\documentclass[12pt]{article}

\newcommand{\be}{\begin{equation}}
\newcommand{\ee}{\end{equation}}
\newcommand{\bea}{\be \begin{array}{rcl}}
\newcommand{\eea}{\end{array}\ee}
\newcommand{\ra}{\rightarrow}

\newcommand{\Sp}{\;\;\;\;}

\newcommand{\tT}{\overline{T}}

\newcommand{\dt}[1]{\dot{#1}}
\newcommand{\ddt}[1]{\ddot{#1}}
\newcommand{\dx}[1]{#1'}
\newcommand{\ddx}[1]{#1''}
\newcommand{\dddx}[1]{#1'''}

\title{Stochastic Quantum Trajectories \\ without a Wave Function}
\author{Jeroen C. Vink\thanks{Jeroen.Vink@Shell.com} \\
Shell Global Solutions (US) Inc. \\
Shell Technology Center Houston - Woodcreek Campus,\\ 200 Dairy Ashford N, Houston, USA.}
\date{December 23, 2014}

\begin{document}
\maketitle

\abstract{After summarizing three versions of trajectory-based quantum mechanics, it is argued
that only the original formulation due to Bohm, which uses the Schr\"odinger wave function to guide the particles,
can be readily extended to particles with spin. To extend the two wave function-free formulations, it is argued
that necessarily particle trajectories not only determine location, but also spin. Since spin values
are discrete, it is natural to revert to a variation of Bohm's pilot wave 
formulation due originally to Bell. 
It is shown that within this formulation with stochastic quantum trajectories, a wave function free 
formulation can be obtained. }

\section{Introduction}
In the last ten years or so, there has been a growing interest
in Bohm's trajectory based interpretation of quantum mechanics \cite{BohmHiley, QuantChem}. 
Even though the introduction
of classical-like particle trajectories into the quantum mechanical state description
does not have any observable consequence, it offers more practical ways to depict and
explore quantum behavior. Specifically, in quantum chemistry this has
led to novel ways to numerically solve and investigate multi-particle quantum systems \cite{QuantChem}.
From a conceptual or ontological point of view, Bohm's interpretation has a number of well-known
advantages; the most compelling feature being the resolution of the measurement problem:
unlike the traditional Copenhagen interpretation, there is no need for a collapse of 
the wave function.
Obviously Bohm's trajectory based interpretation is not generally adopted and one can
raise various objections and queries. For example: Why is there still a wave function 
in this picture - why should the particles need to be guided by a pilot wave? 
Why would there only be trajectories for position (or for fermion number,
if we adopt Bell's version of Bohm's interpretation \cite{Bell})? 
The experimental verification of Bell's inequalities \cite{BellInequalities} has effectively excluded local
hidden variable interpretations of quantum mechanics. The presence of the pilot wave
and its manifestation as a quantum potential are presumed to provide the required
non-locality that allows compliance to Bell's inequalities, but how does this come about?

The first question has been addressed
by Holland, Poirier and Hall et al.~in recent work \cite{Holland05,Poirier10,Hall14}. 
Interestingly, these authors show that Bohm's quantum trajectories can be obtained 
without a guiding wave function from a reformulated theory that prescribes the equations
of motion for the particles, along with a probability distribution for the resulting
particle trajectories. This probability distribution (or, in the version of ref.~\cite{Hall14},
the repulsion among trajectory realizations) generates an additional force that
causes all required quantum effects. 

The second question was addressed in \cite{Vink93}, where it was shown that it is
in fact possible to compute trajectories for all
observables, including inherently discrete entities like spin. In this work
it was argued that one could choose a preferred complete set of (commuting) observables,
or - lacking compelling arguments to prefer one set over another - allow ``all''
observables, also if they are mutually non-commuting, to have well-defined trajectories.

The third question of how non-locality manifests itself in Bohm's quantum trajectory
interpretation, has been discussed in detail in ref.~\cite{BohmHiley}. Here it is shown
how coherence in the multi-particle wave function acts as a non-local guidance for the particle
trajectories and thus manages to avoid the constraints imposed by Bell's inequalities.
Note that the pilot wave, which carries the entangled spin state, is essential 
to make the particles behave such that results of normal quantum mechanics
are reproduced.

This paper revisits the above three points to further clarify the issues
and show their interconnectedness. First, in section \ref{Sect2} three available
formulations of Bohm's trajectory interpretation of quantum mechanics for spin-zero 
particles are summarized. Since all three are designed to reproduce observable results
of normal quantum mechanics, their difference is primarily in the underlying ontology,
or in what are assumed to be ``elements of reality''. Additional differences,
for example how amenable each formulation is to numerical evaluation, will not be 
pursued in any detail here. The first formulation, supported by Bohm \cite{BohmHiley}, 
Bell \cite{Bell} and others \cite{DurrGoldstein} firmly assumes a pilot-wave, which
is a solution of the Schr\"odinger equation, as part of reality; the two other 
formulations make no explicit reference to a wave function or Schr\"odinger 
equation \cite{Holland05,Poirier10,Hall14}. 

Then, in section \ref{Sect3}, it will be argued that such wave function-free formulations
of quantum mechanics that only give a ``beable'' status to position, 
when carried over to particles with spin, will lack the
state information that is required to provide the quantum correlations
required to describe particles with spin. This requires that also spin state
has to be included somehow as an element of reality, as is the case for the
stochastic trajectory formulation of quantum mechanics developed in ref.~\cite{Vink93}. 
Hence, it suggests to explore if the wave function-free formulations discussed 
in section \ref{Sect2} can be extended to this stochastic trajectory formulation.
This is carried out in section \ref{Sect4}, where it is shown that the
ensemble of quantum trajectories for any (discrete) observable can be generated
self consistently from a suitably chosen distribution of initial values by applying a local (stochastic)
evolution rule without reference to a wave function. As in the wave function-free particle
formulation, the dynamics of the stochastically evolving quantum numbers is affected
by the local potential and non-local probability distribution defined by the ensemble of trajectories
in the high-dimensional state space.
Finally, some concluding comments are presented in Section \ref{Sect6}.

\section{Bohm Trajectories without Wave function: Three formulations} \label{Sect2}
To keep notations simple and following refs.~\cite{Poirier10,Hall14}, this section will
consider quantum mechanics of a single particle without spin in one dimension. 
It is mostly straightforward to extend
the discussion to non-relativistic quantum mechanics for multiple spin zero
particles in three space dimensions.
The Schro\"odinger equation for such a one-particle system is given by
\be
 i\hbar\dt{\psi}(x,t) = -\frac{\hbar^2}{2m}\ddx{\psi}(x,t) + V(x)\psi(x,t), \label{schrEq}
 \ee
where $m$ is the particle mass, $V$ the classical potential and $\hbar$ Planck's
constant.
As  explained in e.g.~\cite{BohmHiley}, for a quantum system described by the
wave function $\psi(x,t)$ one can define an ensemble of particle trajectories $x(t, x_0)$.
A family of trajectories can be derived from the wave function 
\be
\psi = Re^{iS/\hbar}. \label{polarPsi}
\ee
using the ``pilot'' equation,
\be
\dt{x}(t,x_0) = \dx{S}(x_0,t)/m, \label{pilotEq}
\ee
were $\dx{S}(x_0,t)$ is the gradient of the phase of the wave function,
evaluated at $x(t_0)=x_0$.
It follows from the Schr\"odinger equation that these particle trajectories 
obey almost classical equations of motion,
\be
\ddt{x} = -\dx{(V(x) + Q(x,t))}/m. \label{eom}
\ee
Quantum effects are due to the quantum potential that is present in addition to the
classical potential $V$,
\be
Q = \frac{-\hbar^2}{2m}\frac{ \ddx{R}}{R}. \label{quantumPotential}
\ee
Quantum uncertainty 
arises because one must consider the full ensemble of trajectories, with all possible
initial conditions $x_0$, which must be distributed  initially according to 
the probability density at starting time $t_0$, $P(x_0,t_0)$. The probability
density for particle positions is computed from the radial part of the wave function as, 
\be
P(x,t) = R^2(x,t). \label{PvsR}
\ee
When evaluating probabilities as averages over this evolving ensemble of particles,  
the same results as in normal quantum mechanics are recovered.

To solve the dynamics of this ensemble of particles, one can follow at least
three approaches, which will be summarized next. 

\subsection{Trajectories Guided by a Wave Function (F-I)} \label{F1}
The most straightforward (and oldest \cite{BohmHiley,DurrGoldstein}) 
approach is to first solve the (linear) Schr\"odinger equation
to obtain the complex valued $\psi(x,t)$.
After computing $\psi(x,t)$, one can use its phase $S(x,t)$ with eq.~(\ref{pilotEq}) 
to compute particle trajectories. The radial part of the wave function does not
play a direct role in this formulation, other than providing the probability
distribution for the initial trajectory positions $x_0$.\footnote{It is even
argued \cite{BohmHiley,DurrGoldstein} that the proper probability distribution, $P(x,t) = R^2(x,t)$  
will develop dynamically from an arbitrary initial distribution, since only this
specific distribution is conserved during the non-linear dynamics of the particles.}

\subsection{Trajectories with Self-contained Dynamics (F-II)} \label{F2}
Second, one can exploit the particle aspect of the dynamics more explicitly
and compute particle trajectories from the equation of motion (\ref{eom}) for 
the particles. Unfortunately, this equation contains the quantum potential, which
is only given at the initial time and subsequently evolves dynamically as well.
In contrast with the previous formulation F-I, the quantum potential 
in this formulation is computed as an emergent property of the ensemble
of particles; The function $R(x,t)$ needed to compute the quantum potential (\ref{quantumPotential})
is obtained as the square root of the probability
density of particle positions in the evolving ensemble of particle trajectories.

Hence, one can apply the following leap-frog scheme for solving the particle
dynamics:
Choose a large set of initial positions, according to a (given) probability
density, $P(x_0,t_0)$, along with initial values for the particle
velocities, $v(x_0, t_0)$.
Compute the quantum potential at this initial time, using (\ref{PvsR}) and (\ref{quantumPotential}).
Make sure that the initial velocities satisfy suitable integrability conditions \cite{Wallstrom94},
such that they can be obtained as the gradient of an underlying scalar function,
\be
v(x,t_0) = \dx{S}(x,t_0)/m. \label{vIsGradient}
\ee
Note that this is not straightforward, nor particularly natural, in more than one dimension.
Then evolve the particle positions in
the ensemble from $t_0$ to $t = t_0 + dt$, using the almost classical equations of motion,
\be
v(x,t+ dt)  = -\frac{1}{m}\dx{(V(x) - 
          \frac{\hbar^2}{ 2m}\frac{\ddx{(P^{1/2}(x,t))}}{ P^{1/2}(x,t)})}dt, \label{eom2a}
\ee
\be
x(x,t+dt)  =  x(x,t) + (v(x,t+dt))dt. \label{eom2b}
\ee
Here, the quantum potential is computed from the probability density
for particle positions, using eq.~(\ref{PvsR}). Notice that the velocity is
updated by adding a gradient, hence condition (\ref{vIsGradient}) remains valid
also for later times.
After propagating all particle positions in the ensemble to the incremented time, 
the new probability density can be computed using a suitable histogramming or averaging technique,
\be
P(x, t+dt) \propto ({\rm \# particles}\in [x,x+a]) / a. \label{sampleP}
\ee
It can be shown \cite{Holland05,Poirier10,Hall14,SchiffPoirier} that this formulation 
generates fields $R(x,t)$ and $S(x,t)$, which can be computed
from $P(x,t)$ and $v(x,t)$, and which can be combined into a wave function 
$\psi(x,t) = R(x,t)e^{iS(x,t)/\hbar}$ that solves the Schr\"odinger equation (\ref{schrEq}).
Hence, as in F-I, this formulation produces the same observable results as normal quantum mechanics.
This is the case when an infinitely large ensemble of trajectories is used to
compute the probability density (\ref{sampleP}). 
When a finite ensemble is used
to estimate the distribution, as proposed in \cite{Hall14}, the quantum potential 
featuring in (\ref{eom2a}) will be an approximation as well, and deviations 
from normal quantum mechanics will be implied.

Even though this scheme looks attractive from a numerical point of view, because it is relatively simple to solve
the time evolution of the particles, the obvious drawback is the difficulty in
computing the time evolution of the quantum potential (or, equivalently, the
probability density of particle positions). One would need either clever
approximation methods, or a very large ensemble of particle trajectories - assuming
that only very small observable deviations from normal quantum mechanics can be allowed. 
In particular when the number of particles increases, the high dimensionality of the 
state space over which the probability density must be evaluated  
will require a very large ensemble size that will make any sampling method quickly prohibitively expensive.

\subsection{Trajectories with Self-consistent Ensemble Dynamics (F-III)} \label{F3}
The third formulation pays tribute to the quantum fluid approach originally introduced
by Madelung \cite{Madelung27} and more recently picked up by the quantum chemistry
community \cite{QuantChem,Poirier10}. It uses the underlying particle dynamics, 
as in section~\ref{F2} above, but use a different method to compute the 
evolving quantum potential. The formulation is best explained in discretized from,
where the underlying space is replaced by a grid. In one dimension, the particle 
positions are then given by 
\be
x_k = ka,  \label{discreteX}
\ee 
where $a$ denotes the grid spacing, which for simplicity is taken to be constant 
throughout space and $k$ is the grid cell index.
Similarly as before, the initial state of the system is defined by specifying
for all $k$, the probability
density $P_k(t_0)$ for the particle to be located in cell $k$, along with the
initial value of the velocity $v_k(t_0)$ that the particle would have in this cell.
Hence, the state of the single particle is fully defined by the values of
$P_k$ and $v_k$ for all $k$. 

As in formulation F-II, the initial values of the velocities in each cell must
be chosen subject to (the discretized version of) the constraint (\ref{vIsGradient}), and
the velocities are updated using the almost classical equations of motion 
(\ref{eom2a}). The probability density is updated by enforcing
the continuity equation,
\be
 P_k(t + dt) = P_k(t) + (P_{k-1}(t)v_{k-1}(t) - P_k(t)v_k(t))dt/a, \label{PvsV}
\ee
where it is assumed that the particles can only move to their direct neighboring
cell, and that the velocity in cell $k$ is in the positive direction.

In this ``quantum hydrodynamics'' formulation as it is presented above, there is no need 
to first solve the Schr\"odinger equation, in order to compute the ensemble of particle 
trajectories. Hence, also this third scheme describes particle dynamics without 
a guiding wave function. However, it is also clear that the computational
effort to simultaneously solve $P_k$ and $v_k$ for all $k$ could be grosso-modo the same as 
solving the the Schr\"odinger equation for $R$ and $S$. 

\subsection{Similarities and Differences}
Before closing this section, it may be worthwhile to summarize similarities
and differences between the three formulations above. It is assumed here that
the ensemble in formulation F-II (section \ref{F2}) is sufficiently large, such 
that the probability density can be accurately computed, and that the grid spacing $a$
in formulation F-III (section \ref{F3}) is sufficiently small to avoid discretization
effects. For a given initial state, all three methods then produce the same trajectories 
for an ensemble of particles and reproduce all observable results of normal quantum mechanics.
Also, as required \cite{BellInequalities}), all formulations have non-locality 
built into their dynamics: in all formulations, particle trajectories are
influenced by the same quantum potential, which is an inseparable function over the multi-particle
state space and hence intrinsically non-local.

There will, of course, be differences in computational strategies that are most natural 
or most efficient when implementing a numerical scheme for each of these formulations. 
However, this is not further pursued here. There are also important differences in the 
ontology implied by each approach, which will be outlined next.

In the traditional Bohm interpretation F-I, the wave function along with a
particle trajectory, are elements of reality. The express purpose of the wave 
function is to guide the particle along its quantum trajectory. 
Quantum effects, such as interference effects but also
entanglement of (spin) states and non-locality, are carried over from the wave 
function, which is evaluated over full state space, to the
quantum potential \cite{BohmHiley} or transition probabilities \cite{Bell},

In formulation F-II, there is a very large, possibly infinite, ensemble of
particle trajectories, but no wave function. Particles move according to their
equations of motion, as in classical dynamics, but the equation of motion are
endowed with an extra potential (or force) term. This (again inseparable) 
quantum potential is generated as a collective effect of the entire ensemble 
of coexisting state space trajectories. Hence, all trajectories are equally 
important elements of reality.

For formulation F-III one can adopt an ontology that is close to either of 
the two above, depending on how the ensemble of trajectories is interpreted.
One can represent the ensemble using the probability density and velocity field
and give these a similar status as (the modulus and phase of) the wave function
in the Bohm interpretation F-I. These evolving probability and
velocity fields produce a quantum potential that, along with the classical potential, 
determines the (single) trajectory for the particle. Hence, almost as in F-I, 
the particle trajectory along with the probability density and velocity field 
are elements of reality. This flavor will be referred to as F-IIIa.
Alternatively, one can represent the ensemble using a (very large but not necessarily
infinite) ensemble of trajectories and effectively adopt the same ontology or
elements of reality as in F-II. This alternative will be referred to as F-IIIb.

\section{Particles with Spin} \label{Sect3}
Until now, the particles were simple point-particles, with only `position' as
intrinsic property that is an element of reality. One can argue \cite{BohmHiley}, 
that it is in fact sufficient that a particle {\it only} has a well-defined position.
Other observable properties, such as momentum or spin, need not be intrinsic properties
of the particle (and hence elements of reality) since they are only indirectly observed though
various types of measurements that in the end boil down to registrations of
pointer positions or other (displayed or printed) position-type indicators. 
Even though a well-defined spin is not an actual or intrinsic property of the moving particle,
its trajectory can still be influenced by spin, because the guiding wave function
evolves differently depending on its spin attributes, which are driven by
spin-dependent terms in the Hamiltonian. In this way the results of the
measurement process involving particles spin, including particles with entangled states,
can reproduce the normal results of quantum mechanics \cite{BohmHiley}[ch.10,12],\cite{Bell}[Ch.18]. 

Since wave function-free formulations require that the dynamics of the evolving
state can be computed self-consistently without recourse to a guiding wave function,
this seems to preclude extending formulations F-II and F-III to quantum mechanics
for particles with spin, unless the particles are also endowed with spin degrees of freedom.
As Bohm noted \cite{BohmHiley}[Ch.10.4], this appears to be difficult, since spin degrees of freedom linked to a 
particle would  grow proportionally with the number of particles, whereas the actual number 
of spin degrees of freedom grows exponentially. Hence, Bohm conceded that for particles with 
spin, a formulation in which trajectories are determined by the influence of a combination 
of classical and quantum forces could no longer be maintained
If it is possible to avoid Bohm's conclusion and find a wave function-free 
formulation for trajectories for particles with spin, it appears inevitable to 
start from an ontology that includes full multi-particle spin state
as an element or reality. Such a formulation was presented in ref.~\cite{Vink93},
which showed how ``de Broglie-Bohm-Bell'' (BBB) trajectories could be computed for any 
observable - also for multi-particle systems with both position and spin. 
The next section will show how this approach can be reformulated to provide a wave 
function-free description of the dynamics.

\section{BBB Trajectories without Wave Function} \label{Sect4}
Following ref.~\cite{Vink93}, all observables are assumed to take discrete
values in this section. 
The Schr\"odinger wave equation for any discrete quantum system, including multi-particle systems with
with spin can be written as
\be
 i\hbar \dt\psi_{n} = \sum_m H_{nm}\psi_{m}, \label{genericWaveEq}
\ee
where the composite index $n$ labels the discrete values of the relevant
observables. For example, eq.~(\ref{genericWaveEq}) can represent the Pauli equation for
two-component wave functions and particles moving on a line or circle. Then
$n \equiv k,s$, with  $k = x_k/a$ labeling position (cf.~(\ref{discreteX})), and $s$ 
labeling the spin state.

Stochastic trajectories for the discrete values of the system state
$n$ can be computed as follows. In a small time interval $dt$, state $m$ will jump to
$n$ with probability $P(m\ra n)$, where
\be
P(m \ra n) = T_{nm} dt.
\ee
The transition probability $T_{nm}$ is defined as
\be
T_{nm} = \max(0,J_{nm}/P_m), \label{TfromJ}
\ee
where $J$ is the probability current,
\be
J_{nm}=\frac{2}{\hbar}{\rm Im}(\psi^*_n H_{nm}\psi_m), \label{defJ}
\ee
which is anti-symmetric in $n$ and $m$, 
and $P_m$ is the usual probability for state $m$,
\be
P_m = \psi^*_m\psi_m.
\ee
The matrix $H_{mn}$ is the Hamiltonian in the $n$ representation. Further details
are in refs.~\cite{Vink93} and \cite{Bell}[Ch.19].
The above prescription for $n$-trajectories has the form of a guidance relation
similar to (\ref{pilotEq}) for (continuous) particle position: to compute
the transition probabilities for time $t$, the wave function $\psi_n$ at time
$t$ must be computed from the wave equation (\ref{genericWaveEq}). 

To obtain a wave function-free dynamics, the time-dependence of $T_{nm}$ 
should be computed without recourse to a guiding wave function. As in
formulation F-III, the initial state should then be defined as an ensemble
of $n$-values, with prescribed probabilities $P_n$. The stochastic trajectories along with
the evolving probabilities $P_n(t)$ can then be computed as,
\be
\dt{P}_n(t) = P_n(t) \sum_m(T_{nm}(t)P_m(t) - T_{mn}(t)P_m(t)) 
\ee
provided that the time-dependence of $T_{nm}$ can be computed from a relation 
like
\be
 \dt{T}_{nm}(t) = F(T,P). \label{Tdot}
 \ee
The function $F$ may depend on the transition rates $T$ and probabilities $P$,
but must not refer to a pre-computed wave function as in (\ref{TfromJ},\ref{defJ}).

To discover the evolution equation (\ref{Tdot}), it is convenient to first find an
evolution equation for the currents $J_{nm}$, and then use (\ref{TfromJ}) to
convert it into an evolution equation for the transition rates $T_{nm}$.
The first step  then is to rewrite the probability
current(\ref{defJ}) using a wave function in polar form (\ref{polarPsi}),
\be
J_{nm} = -\frac{i}{\hbar}(R_n H_{nm} R_m e^{-i(S_n-S_m)/\hbar} - R_m H_{mn} R_n e^{i(S_n-S_m)/\hbar})
         \label{polarJ}
\ee
and use this relation to solve for the wave function phase, or rather for 
\be
  \theta_{nm} = e^{-i(S_n-S_m)/\hbar}.
  \ee
Since (\ref{polarJ}) leads to a quadratic equation for $\theta_{nm}$, there will
be two solutions for a given $J_{nm}$. However, consistency requires
that $\theta_{nm} = 1$ when $S_n = S_m$, which implies that the solution takes the form, 
\be
\theta_{nm} = \frac{[H_{nm}]\alpha_{nm} + i\hbar J_{nm}}{2R_m R_n H_{nm}} ,
                        \label{theta1}
\ee
where
\be
[H_{nm}] = +{\rm Sign}({\rm Re}(H_{nm})) |H_{nm}|   \label{signFix}
\ee
and
\be
\alpha_{nm} = (4 R_m^2 R_n^2 - \frac{\hbar^2 J^2_{nm}}{|H_{nm}|^2})^{1/2}. \label{alpha}
\ee 
It can be readily shown that the term in the square root is non-negative, hence
$\alpha$ is real valued, with $\alpha_{nm}=\alpha_{mn} \ge 0$.
The definition (\ref{signFix}) shows how the sign ambiguity is fixed; 
note that $[H_{mn}]=[H_{nm}]$ and $[H_{nm}] = H_{nm}$ for a real valued Hamiltonian matrix.

As stated above, a wave function-free dynamics, requires an expression for $\dt J$, which 
may depend on $J$, $R$ (or $P$) and the Hamiltonian, but not on the phases $S$. 
In such a formulation, the phases $S$ (or equivalently, the $\theta$) are derived
properties, which can be back-computed from the evolving values of $J$ and $R$. 
It should be noted that the initial values for $T$, or equivalently $J$, 
cannot be chosen freely, since $J_{nm}$ must obey (\ref{polarJ}) in which the $R_n$
and $S_n$ are the independent degrees of freedom\footnote{It is assumed here 
that there are enough non-zero off-diagonal matrix
elements in the Hamiltonian, such that for each independent $S_n$ (except one, because
the overall phase is irrelevant) there is at least one non-zero element
$H_{nm}$, and hence $J_{nm}$.}. 

The sign-choice implemented above assumes that
in the initial state the real part of $\theta_{nm}$ is larger than zero. 
If, during subsequent time evolution 
\be
\alpha_{nm}\ra 0,   \label{crossover}
\ee
it may be required to flip the sign in (\ref{signFix}) such that $\theta_{nm}$
has a continuous time dependence.

Using the result (\ref{theta1}) the time-dependence of $J_{nm}$  can
be computed from
\be
 \dt{J}_{nm} = \frac{2}{\hbar^2}{\rm Re}(\sum_{k|n}R_k H_{kn}H_{nm}R_m\theta^*_{mn}\theta^*_{nk}
            - \sum_{k|m}R_n H_{nm}H_{mk}R_k\theta_{nm}\theta_{mk}) 
\ee
as
\be
\dt{J}_{nm} = \frac{1}{2 R_n^2}(J_{nm}\sum_{k|n}J_{nk}
           -\frac{1}{\hbar^2}[H_{nm}]\alpha_{nm} \sum_{k|n}[H_{nk}]\alpha_{nk})
           - ( m \leftrightarrow n ). \label{dotJ}
\ee
The notation $k|n$ in the summations indicates that only $k$ values for which
$H_{kn} = H^*_{nk}$ is not equal to zero are included in the sum.

To turn eq.~(\ref{dotJ}) into an evolution equation for the transition rates
it is convenient to introduce an generalized transition rate $\tT$, defined as
\be
\tT_{nm} = J_{nm}/P_m, \label{tildeTfromJ}
\ee
where $\tT_{nm}$ is allowed to be negative, and
\be
  \dt{\tT}_{nm} = \dt{J}_{nm}/P_m - (J_{nm}/P_m) \dt{P}_m / P_m.
\ee
Using (\ref{dotJ}) and the continuity equation
\be
  \dt{P}_m = \sum_{k|m}J_{mk},
\ee
the final result is obtained as,
\bea
\dt{\tT}_{nm}& = & -\frac{1}{2}\tT_{nm}(\sum_{k|n}\tT_{kn} - \sum_{k|m}\tT_{km}) \nonumber \\[5mm]
           & + & \frac{1}{2\hbar^2}P^{-1}_m[H_{nm}]\alpha_{nm}( P^{-1}_n\sum_{k|n}[H_{kn}]\alpha_{kn}
                                                    -  P^{-1}_m\sum_{k|m}[H_{km}]\alpha_{km} ) 
    \label{tildeTfinal}
\eea

With this result, a wave function-free formulation for the stochastic
$n$-trajectories has been obtained, which is similar to formulation F-III. Specifically,
the procedure to compute stochastic trajectories for the $n$-values is as follows.
Define initial conditions for the trajectories of all $n$, by choosing
probabilities $P_n$ and transition rates $T_{nm}$. Since not all $T_{nm}$ can be
chosen independently, it is actually easier to choose initial values for the
phase $S_n$ and then use eqs.~(\ref{polarJ}) -- along with the chosen values for
$R_n=P_n^{1/2}$ -- to compute the $J_{mn}$ and from these
the $\tT_{mn}$ and $T_{mn}$. In this way $T_{mn}$ will satisfy all required consistency conditions and
subsequent evolution using (\ref{updatePT}) will preserve these conditions.
Finally, a sufficiently small time step
size $dt$ must be chosen, such that also $T_{mm}dt = 1 - \sum_n T_{nm}dt > 0$ holds for all $m$.
The time-dependence of the system is obtained by  iteratively updating the 
probabilities $P_n$ and transition rates $T_{nm}$, using
\bea
P_n(t+dt) & = & P_n(t) + \sum_m( T_{nm}(t)P_m(t) - T_{mn}(t)P_n(t)), \label{updatePT} \\[4mm]

\tT_{nm}(t+dt) &  = & \tT_{nm}(t) + \dt{\tT}_{nm}(t) dt, \\[4mm]

T_{nm}(t+dt) &  = & \max(0, \tT_{nm}(t+dt)),
\eea
where $\dt{\tT}$ is computed from eq.~\ref{tildeTfinal}).
The sign-choice implied in the $[H]$-terms in (\ref{tildeTfinal}) should be the 
same as used for the previously time step, unless the cross-over condition 
(\ref{crossover}) is encountered, which might required to flip the sign in (\ref{signFix}).

\subsection{Particle Moving on a Circle}
The update algorithm (\ref{updatePT}) is conceptually similar to 
formulation F-III for spin-zero particles in continuous (1D) space.
To show this more explicitly, it is instructive to compute the update
rules in the continuum limit for a particle moving on a circle,
for which the Hamiltonian can be written as (cf.~\cite{Vink93}),
\be
H_{mn} = (V_n + \frac{\hbar^2}{Ma^2})\delta_{m,n} 
               - \frac{\hbar^2}{2Ma^2}(\delta_{m+1,n}+\delta_{m,n+1}),
               \label{particle1D}
\ee
with $V_n$ the potential, $M$ the particle mass and $a$ the grid spacing.
In this case, the evolution equations of formulation F-III should be recovered.

For a state in which the ``particle'' moves in the positive $n$ direction, it is
sufficient to compute the transitions rates from $n$ to $n+1$. 
This rate multiplied with the grid cell spacing $a$,
can be interpreted as the average particle velocity, $v_n$ (cf.~(\ref{PvsV})). 
Using this change in notation, the generalized transition rates can be written as
\be
 \tT_{n+1,n} = \dt{v}_n /a, \Sp\Sp \tT_{n,n+1} = -v_n P_n/aP_{n+1}.
\ee
Using this definition of velocity, the evolution equation (\ref{tildeTfinal}) turns into
\be
\dt{v}_n = \frac{1}{2a} v_n( -v_{n-1}\frac{P_{n-1}}{P_n} + v_n 
                           + v_n\frac{P_n}{P_{n+1}} - v_{n+1})  \label{vdot}
    + ({\rm \mbox{H-terms}}),
\ee
where ``H-terms'' indicate the second group of terms in (\ref{tildeTfinal}), which will
be considered shortly.
Since the particle moves in one space dimension, the summations in (\ref{tildeTfinal})
only have two terms, which are explicitly shown in (\ref{vdot}).
To investigate the continuum limit, $a \ra 0$, the velocities are expanded to
order $a$, $v_{n \pm 1}=v_n \pm a\dx{v}_n$ and similarly for $P_{n\pm 1}$. Substitution
in (\ref{vdot}) shows that $O(1)$ and $O(a)$ terms cancel, such that only
the H-terms remain. 

Eq.~(\ref{particle1D}) shows that $H_{n+1,n} = O(1/a^2)$, from which it follows that 
$\alpha_{n+1,n}=2R_{n+1}R_n + O(a^2)$ and hence,
\be
 \dt{v}_n = \frac{2a}{\hbar^2}\frac{[H_{n+1,n}]}{R_n^2}
        ( R_n\sum_{k|n+1}R_k[H_{k,n+1}] - R_{n+1} \sum_{k|n}R_k[H_{k,n}] )+ O(a).
                        \label{vdotR2}
\ee
Since the Hamiltonian matrix elements are real valued, the square brackets can
be dropped and (\ref{particle1D}) implies that
\be
[H_{n+1,n}]=[H_{n,n+1}] = -\hbar^2/2Ma^2, \Sp\Sp [H_{n,n}] = V_n + \hbar^2/Ma^2.
\ee
After substituting this in (\ref{vdotR2}) and expanding the shifted terms 
$R_{n+k}$ to third order in $a$ as $R_{n+k}=R_n + ak\dx{R}_n + a^2k^2\ddx{R}_n/2
   + a^3k^3\dddx{R}_n/6$, one finds that $O(1)$, $O(a)$ and $O(a^2)$ terms 
cancel and the expected result emerges,
\be
  \dt{v}_n = -\frac{1}{M}(V_n - \frac{\hbar^2 \ddx{R}_n}{2MR_n})' + O(a).
\ee

\subsection{Particle with Spin 1/2 }
To keep this example simple, the particle is assumed to have a very large mass
such that its motion can be ignored, and to only interact with a magnetic field
$B$. Hence, the $2 \times 2$ Hamiltonian takes the simple form,
\be
  H = \mu \sigma \dot {\mathbf B} = \mu B \sigma_x, \Sp\Sp
  \sigma_x = \left( \begin{array}{cc}
         0 & 1 \\
         1 & 0 
  \end{array}\right),
\ee
where the $x$ axis has been aligned with the magnetic field.
Since $n$ takes only two values, there is only one independent phase
(the arbitrary global phase can be used to set one of the phases equal to 0) and
one independent probability (since $P_1 = 1 - P_2$). Therefore there is also 
only one transition rate $\tT_{12}= -\tT_{21} P_1/P_2$
that has to be solved. Eq.~(\ref{tildeTfinal}) therefore simplifies to
\be
 \dt{\tT}_{12} = \frac{1}{2}\tT_{12}^2(1+\frac{P_2}{P_1}) +
    \frac{\alpha_{12}^2[H_{12}]^2}{2\hbar^2 P_2}(\frac{1}{P_1}-\frac{1}{P_2}).
\ee
Using $[H_{12}] = \mu B$ and
\be
\alpha_{12} = (4P_1 P_2 - \frac{\hbar^2}{\mu^2 B^2}\tT_{12}P_2^2)^{1/2}
\ee
it follows that
\be
\dt{\tT}_{12} = \tT_{12}^2 + \frac{2\mu^2 B^2}{\hbar^2}(2 -\frac{1}{P_2}).
     \label{dotTspin}
\ee
Differentiating the continuity equation $\dt{P}_2=-\tT_{12}P_2$ and eliminating
$\dt{\tT}_{12}$ using (\ref{dotTspin}) gives
\be
  \ddot{P}_2 = -2\frac{2\mu^2 B^2}{\hbar^2}(P_2 - 1/2).
\ee
This equation is easily solved as
\be
P_2(t) = (1 + \cos( 2\gamma t + \delta))/2 =\cos^2(\gamma t + \delta),
\ee
with $\gamma = 2\mu^2 B^2/\hbar^2$ and $\delta$ a free integration constant.
The generalized transition rate then follows as
\be
\tT_{12} = 2\gamma \tan(\gamma t + \delta).
\ee
It is straightforward to check that these results exactly agree with the 
results obtained from the wave function that solves the Schr\"odinger equation
for this system.
  
\section{Revisiting Entangled Spin States} \label{Sect5}
With the wave function free formulation as developed above, it is clear that also
quantum non-locality and entangled spin states can be properly captured. This is the
case, because there is a one-to-one relation between the dynamics of the
probability field $P_n$ and (generalized) transition probabilities $T_{nm}$ and
the modulus $R_n$ and wave function phases $S_n$ - provided that the initial
transition rates (or equivalently currents $J_{nm}$ are chosen in accordance 
with the constraint (\ref{polarJ}).
Hence, the description of the measurement process, including non-local effects
due to entangled quantum states that was presented in \cite{BohmHiley}[Ch.6,7] 
remains valid for the alternative formulation presented here. 
In formulation F-I the guiding wave function contains the
extra spin state information, in the stochastic wave function-free dynamics,
the state itself, which includes spin and position values as well as velocities
and transition rates, contains all required information. Even though the
evolution equations (\ref{updatePT}) may seem to be local, they of course describe
the dynamics of a single point in the high-dimensional state space. An ensemble
of such states can capture all required \cite{BellInequalities} correlations between 
the individual components of this state vector, also when these components refer to
particle locations and spins at widely separated locations.

\section{Discussion} \label{Sect6}
The results of this paper extend previously developed formulations of quantum dynamics
without wave functions to (non-relativistic) particles with spin. Specifically the
formulation developed in section~\ref{Sect3} shows that the stochastic trajectories of 
the alternative formulation of quantum mechanics in terms of discrete ``beables'' \cite{Vink93}, 
can be generated without recourse to a guiding wave function.
Exploring such an alternative might offer new insights. In particular it simplifies
the ``beable'' world, or ontology, compared to a world in the Bohm-Bell formulation, F-I, which
needs two different entities, particle locations with classical-like trajectories
and a complex valued wave function that evolves according to the Schr\"odinger equation.
Also, having an alternative set of evolution equations could lead to different,
possibly more efficient numerical methods to solve and depict quantum dynamics.

However, the stochastic dynamics developed above also has less attractive features.
Since the dynamics is computed in terms of the over-complete set of generalized transition rates,
instead of the wave function phases, the initial conditions are constrained in a
rather complex and unnatural fashion.
Something similar is present in the wave function-free formulations F-II and F-III: here the
particle trajectories have a constraint on the initial velocities (two or more dimensions).
Even formulation F-I has a constraint on the initial probability density for particle
positions, which must be chosen to be equal to the modulus squared of the wave function. 
However, here Bohm and others \cite{BohmHiley,DurrGoldstein} argue that these constrained initial conditions may not
be required, since deviations will dynamically average out.
In the current formulation, which lacks a guiding wave function, such a dynamical
recovery of the interdependency of probability current and wave function phase
seems very unlikely.

Also the gain in simplicity compared to the Bohm-Bell formulation F-I is perhaps
fairly meager: As in formulation F-III, one still needs to solve for the 
probabilities and transition rates in the full state space, i.e, an equally imposing
task as solving the guiding wave function in F-I.

Finally, there is the question if it is natural to adopt an integer spin value
as an element of reality. In itself this is attractive, since it avoids having
to argue that some quantum state properties are fundamentally different from
others. I.e., one needs not argue that ``location'' is an intrinsic property, also in the microscopic
world, but ``spin'' is not, and -- unlike position -- only emerges when macroscopic measurements are
performed. Of course, as discussed in ref.~\cite{Vink93}, if spin would be
granted a ``beable'' status, it is still not clear {\it which} spin: In the
formulation developed above, eigenvalues of any spin operator can be ``beables''.
Choosing for example eigenvalues of $\sigma_z$, the probabilities and transition
rates are directly relevant for the evolution of the spin in this $z$-direction.
Measurement result of $\sigma_z$ can then be directly understood, i.e., ``read off'' from the quantum
mechanical state -- as is the case for properties in a classical state. However,
results of measurements of other (non-commuting) spins can no longer be understood
directly and one has to invoke the same arguments as used in F-I (\cite{BohmHiley}[Ch.6,7]
to understand how measurements of these spins will still agree with the results
of normal quantum mechanics.

It may be interesting to note at this point that one could resort to the view proposed
in ref.~\cite{Vink93}, that in fact {\it all} observables have a ``beable'' status.
For spin this would imply that the particle has many coexisting spin properties, 
for all possible spin operators (i.e., a dense, but still finite subset of all possible
Hermitian and unitary $2\times 2$ matrices). Quantum entanglement
then plays out in a perhaps surprising, but very compelling way: In the typical set-up of two
free moving particles with entangled spins, the spin part of the Hamiltonian is zero, and hence
the spin state remains unchanged while the particles move away from each other.
Therefore it may seem that this system has the same (non)locality as a classical system
in which some properties are correlated over long distance, simply because they
started out that way. In order then to understand how the correct quantum results are obtained when
performing spin measurements in arbitrary relative directions -- without evoking
the decohering effect of a measuring apparatus - one must assume that the 
specific set-up of the measuring apparatus, specifically reads-off
the property from the trajectory with a spin direction that aligns with the
device. However, this cannot be done in a local fashion: the specific two particles spin realization
is picked-up for which each of the {\it two} spin operators aligns with the corresponding measuring
device. From a quantum trajectory point of view, the combined measuring apparatus reads-off
a specific state from one trajectory, but this trajectory of course describes the 
correlated spin values for two widely separated particles.

On the one hand, this coexisting states interpretation of quantum mechanics 
compellingly shows the non-local nature of ``The Undivided Universe'';
on the other hand -- to quote Bell -- ``It seems an eccentric way to make a world''.

\end{document}